\documentstyle[12pt]{ioplppt}
\begin{document}
\jl{1}
%
\title{A renormalization group study of a class of 
reaction-diffusion model, with 
particles input}[Renormalization group study of a reaction-diffusion model]
\author{Pierre-Antoine Rey and Michel Droz}
\address{D\'epartement de Physique Th\'eorique, Universit\'e de Gen\`eve,
CH-1211 Geneva 4, Switzerland}
\author{{\bf UGVA-DPT 1996/08-936}}

\begin{abstract}
We study a class of reaction-diffusion model extrapolating continuously between
the pure coagulation-diffusion case ($A+A\to A$) and the pure
annihilation-diffusion one ($A+A\to\emptyset$) with particles input
($\emptyset\to A$) at a rate $J$.  For dimension $d\leq 2$, the dynamics
strongly depends on the fluctuations while, for $d >2$, the behaviour is
mean-field like.  The models are mapped onto a field theory which properties are
studied in a renormalization group approach.  Simple relations are found between
the time-dependent correlation functions of the different models of the class.
For the pure coagulation-diffusion model the time-dependent density is found to
be of the form $c(t,J,D)=(\frac{J}{D})^{1/\!\delta}{\cal
F}[(J/\!D)^{\Delta}Dt]$, where $D$ is the diffusion constant.  The critical
exponent $\delta$ and $\Delta$ are computed to all orders in $\epsilon=2-d$,
where $d$ is the dimension of the system, while the scaling function $\cal F$ is
computed to second order in $\epsilon$.  For the one-dimensional case an exact
analytical solution is provided which predictions are compared with the results
of the renormalization group approach for $\epsilon=1$.

\end{abstract}

\maketitle

\section{Introduction}

It is by now well established that reaction-diffusion models can have a rich
dynamics governed in low dimensions by the fluctuations.  Several physical
quantities behave as power law and the associated critical exponents have some
universal properties.  The renormalization group (RG) method developed in the
framework of equilibrium statistical physics~\cite{Wilson}, provides a suitable
tool to study such dynamics.  To be able to approach the problem of
reaction-diffusion in term of RG, one has first to go from the initial
microscopic master equation to a coarsed-grained description.  The standard way
to do this consists in using a Fock space formalism (see the works of
Doi~\cite{Doi}, Grassberger and Scheunert~\cite{Grass} and
Peliti~\cite{PelOne}).  One ends up with a model which dynamics is defined by
the action of a continuous field theory.  Among several applications, this
method has been used to study the two species annihilation reaction problem
$A+B\to \emptyset$; for homogeneous initial state with equal
densities~\cite{LeeCar}, or unequal densities~\cite{ReyDroz}, rigorous
predictions of Bramson and Lebowitz~\cite{Bramson} have been reproduced and
novel results obtained.  For the case, in which the two species are initially
spatially separated, Howard and Cardy~\cite{Howard}, have confirmed scaling
arguments developed by Cornell and Droz~\cite{Cornell}.

Another interesting family of reaction-diffusion processes is formed by the
one-species diffusion-annihilation and the diffusion-coagulation models.  In
1986, Peliti~\cite{PelTwo} showed that the coagulation-diffusion model $A+A\to
A$ belongs to the same universality class than the annihilation-diffusion one
$A+A\to \emptyset$.  He also showed that the associated field theory is
super-renormalizable.  Thus, only the coupling constant needs to be
renormalized.  Moreover, this renormalization can be done to all orders in
$\epsilon=d_{\rm u}-d$, where $d_{\rm u}$, the upper critical dimension is 2 in
this case.  Peliti showed that the concentration $c(t)$ of the reactant in the
long time regime behaves as:

\[
c(t) \propto t^{-\alpha},\qquad
\alpha=\frac{d}{2},\qquad (t\to\infty).
\]
However, he made no predictions concerning the amplitude, neglecting the initial
conditions in its approach.  It turns out that the initial conditions may play a
very important role.  This aspect has been taken into account within this
formalism only in 1991 by Ohtsuki~\cite{Ohtsuki} and in 1992 by Friedman \etal
\cite{Fried}.  In 1994, Lee~\cite{Lee} gave the first complete RG analysis of
the annihilation-diffusion model.  It turns out that the initial conditions show
up as a local source into the action.  Lee was able to treat this term to all
orders in perturbation theory.  It was shown later (see~\cite{Howard} for the
two-species annihilation and~\cite{ReyUnp} for the one-species case) that such
infinite resummation is equivalent to a shift of the fields in the action by
their classical values.

Meanwhile, in 1993, Droz and Sasv\'ari~\cite{DrozSas} addressed the problem of
both annihilation-diffusion and coagulation-diffusion in presence of a source
$J$ of particles:
\[
A+A\mathrel{\mathop\to^{k}}\emptyset,\qquad
A+A\mathrel{\mathop\to^{g}} A,\qquad
\emptyset\mathrel{\mathop\to^J} A.
\]
Performing a renormalization procedure, they found that the density of particles 
obeys  the scaling law
\begin{equation}
c(t,J,D)
= \biggl(\frac{J}{D}\biggr)^{d/(d+2)}
  {\cal F}[(J/\!D)^{2/(d+2)}Dt],\label{eq:scaling}
\end{equation}
for sufficiently small value of $J$. However, the scaling function $\cal F$ was 
not computed. Making ad hoc assumptions on the asymptotic behavior of $\cal F$ 
in the limits  $t \to \infty$, they showed that the stationary particle density
\[
c\sim J^{1/\!\delta},\qquad \delta=1+\frac{2}{d},
\]
was approached with a characteristic relaxation time $\tau$ given by
\[
\tau\sim J^{-\Delta},\qquad \Delta=\frac{2}{d+2}.
\]
Moreover, considerations on the $J \to 0$ limit, allow them to reproduce the
scaling laws postulated phenomenologically by R\'acz~\cite{Racz}, namely:
\[
\alpha\delta\Delta=1,\qquad \Delta+\frac{1}{\delta}=1.
\]

The goal of this paper is two folds.  First to provide a complete
renormalization group analysis of such models by computing not only the
exponents but also the scaling function $\cal F$ defined by~\eref{eq:scaling} in
the framework of an $\epsilon=2-d$ expansion.  This is done in section 2.
Second, to give an exact analytical solution of the one-dimensional
coagulation-diffusion model with infinite reaction rate and source, by extending
to time dependent regime an approach developed by Doering and
ben-Avraham~\cite{Doeben}.  These exact results are compared with the RG
predictions in the limit $\epsilon=1$ in section 3.  Remarks and conclusions are
given in section 4.

\section{Field theoretical approach and renormalization group analysis}

\subsection{The model and the associated field theory}

We shall not derive here in details how one obtains the field theoretical model.
The interested reader is referred to the original papers of Doi~\cite{Doi},
Grassberger and Scheunert~\cite{Grass} and Peliti~\cite{PelOne,PelTwo} or to the
short reviews presented in~\cite{Lee} and in~\cite{Balb}.  As the number of
particles is not conserved, the basic idea is to introduc a Fock space
representation.  The time evolution operator of the problem can then be cast in
a path integral form which, in the continuous (coarsed grained) limit is
characterized by the action:
\begin{equation}
S_\gamma[a,\bar{a},J]=\int\!\d^{d}x\!\int_{0}^{t}\!\d\bar{t}\,
\biggl[\bar{a}\biggl( \frac{\partial}{\partial\bar{t}} - D\nabla^2\biggr) a
      + \gamma\lambda\bar{a}a^2 + \lambda\bar{a}^2a^2 - J\bar{a}
\biggr], \label{eq:alphaction}
\end{equation}
A whole class of models indexed by the parameter $\gamma\in[1,2]$ is thus
defined.  For $\gamma=1$ one has the pure coagulation-diffusion model and for
$\gamma=2$ the pure annihilation one.  For $1<\gamma<2$, both reactions are
possible with a given probability depending on $\gamma$ (see~\cite{Balb} for
more details).  The coupling constant $\lambda$ is related to the reaction rates
$g$ and $k$ via $\lambda=(\tilde\gamma+1)k$ and
$\gamma=(\tilde\gamma+2)(\tilde\gamma+1)^{-1}$, with $\tilde\gamma=k/g$.  The
particles diffuse in an infinite $d$-dimensional space with a diffusion constant
$D$.  The above action could models two different types of colliding 
particles in some appropriate limits.
First point-like particles living on a $d$-dimensional hypercubic
lattice~\cite{Lee}, second extended particles living in a $d$-dimensional
continuous space~\cite{Balb}.

The time and position dependent fields $a$ and $\bar{a}$ obey bosonic like
commutation relations.  The field $a(x,t)$ is related to the local particle
density, while the auxiliary field $\bar{a}(x,t)$ has no particular 
physical meaning.

Within this formalism, correlation functions are expressed by functional 
integrals.
\begin{eqnarray}\fl
G_{\gamma}^{N,\bar{N}}(\{x_i,t_i\}_1^{N+\bar{N}})
= \int\! {\cal D}a\, {\cal D}\bar{a}\, a(x_1,t_1) \ldots a(x_N,t_N) \nonumber \\
\times
  \bar{a}(x_{N+1},t_{N+1}) \ldots \bar{a}(x_{N+\bar{N}},t_{N+\bar{N}}) 
  \exp(-S_{\alpha}[a,\bar{a},J]). \label{eq:defcor}
\end{eqnarray}
In particular, the particle density $c_{\gamma}(x,t)$ at point $x$ and time $t$ 
reads:
\begin{equation}
c_{\gamma}(x,t)
= \int\! {\cal D}a\, {\cal D}\bar{a}\, a(x,t) \exp(-S_{\gamma}[a,\bar{a},J])
\label{eq:density}
\end{equation}
As we have shown in a previous work~\cite{Balb}, the correlation functions of 
different $\gamma$ models are closely related. In particular, one has
\[
S_{\gamma}[a,\bar{a},J]=S_{1}[\gamma a,\gamma^{-1}\bar{a},\gamma J],
\]
which implies for the concentration
\begin{equation}
c_{\gamma}(x,t;J)=\gamma^{-1}c_{1}(x,t;\gamma J). \label{eq:densrel}
\end{equation}

Accordingly, it suffices to study one particular model belonging to the class to
know the behaviour of the other members.  From now on, we shall study the pure
coagulation-diffusion model ($\gamma=1$), with an initial state empty, which
action is:
\begin{equation}
S[a,\bar{a},J]=\int\!\d^{d}x\!\int_{0}^{t}\!\d\bar{t}\,
\biggl[\bar{a}\biggl( \frac{\partial}{\partial\bar{t}} - D\nabla^2\biggr) a
      + \lambda\bar{a}a^2 + \lambda\bar{a}^2a^2 - J\bar{a}
\biggr], \label{eq:action}
\end{equation}
For the sake of simplicity, we do not write the index $\gamma=1$ in the
following.  When $J=\lambda=0$, one has a free theory (pure diffusion) and the
spatial Fourier transform of the free propagator is simply $G_0(p,t)= \theta(t)
\exp(-Dp^2t)$, where $\theta(t)$ is the usual Heaviside function.  Simple power
counting shows that the upper critical dimension of action~\eref{eq:action} is
$d_{\rm u}=2$.  For $d>d_{\rm u}$, the quadrivertex $\lambda\bar{a}^2a^2$ is
irrelevant and can be discarded.  Below $d_{\rm u}$, the quadrivertex
$\lambda\bar{a}^2a^2$ is relevant and leads to singularities that have to be
renormalized.  At $d=d_{\rm u}$, this vertex is marginal, and one expects
logarithmic corrections to the mean-field behaviour.

\subsection{Mean-field solution}

We first consider the case $d>d_{\rm u}=2$, where the behavior is mean-field
like.  At the mean-field level, the equations of motion for $a$ and $\bar a$ are
obtained from the action~\eref{eq:action}, by the usual saddle point argument
and read:
\begin{equation}
\frac{\delta S}{\delta\bar{a}} 
= \biggl( \frac{\partial}{\partial t} - D\nabla^2\biggr) a + \lambda a^2 
  + 2\lambda\bar{a}a^2 - J
= 0 \label{eq:mf}
\end{equation}
and
\begin{equation}
\frac{\delta S}{\delta {a}} 
= -\biggl( \frac{\partial}{\partial t} + D\nabla^2\biggr) \bar{a}
   + 2\lambda\bar{a}a + 2\lambda\bar{a}^2a
= 0.
\end{equation}
Assuming that $a$ and $\bar{a}$ are homogeneous, it follows that as expected  
$\bar{a}=0$ is a solution and \eref{eq:mf} becomes
\[
\frac{\partial a}{\partial t} = - \lambda a^2 + J
\]
with the initial condition $a|_{t=0}=0$. Thus, the mean-field or classical 
solution is:
\begin{equation}
a_{\rm{cl}}(t) 
= \sqrt{\frac{J}{\lambda}}
  \Biggl(1 -
        2\frac{\exp(-2\sqrt{J\lambda}\,t)}{1+\exp(-2\sqrt{J\lambda}\,t)}
  \Biggr), \label{eq:MFdens}
\end{equation}
where the subscript "cl" stands for classical.

\subsection{Renormalization}

Let us now consider the problem below 2 dimensions.  A brute force computation
of the particle density from~\eref{eq:density} leads to divergences.  Thus the
coupling constant $\lambda$ needs to be renormalized.  Neither the fields $a$
and $\bar{a}$ nor the diffusion constant $D$ require a
renormalization~\cite{PelTwo,Lee}, and as a consequence, the particles input
rate $J$ does not either.

We define the temporally extended vertex function $\lambda(p,t)$ to be the sum
of the diagrams shown in~\fref{fig:one}.  These diagrams can be summed to all
orders and the Laplace-transformed vertex function reads (see~\cite{Lee}):
\begin{equation}
\tilde{\lambda}(p,s)
=\frac{\lambda}
      {1+2[\lambda/(8\pi D)^{d/2}]\Gamma(\frac{1}{2}\epsilon)
       (s + \frac{1}{2}Dp^2)^{-\epsilon/2}} \label{eq:lambda}
\end{equation}
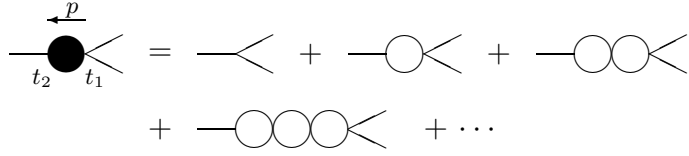
\begin{figure}[tb]
\setlength{\unitlength}{0.5mm}
\centerline{
\begin{picture}(180,40)(0,-20)
\put(0,0){\line(1,0){10}}
\put(15,0){\circle*{10}}
\put(20,0){\line(2,1){10}}
\put(20,0){\line(2,-1){10}}
\put(40,0){\makebox(0,0){=}}
\put(50,0){\line(1,0){10}}
\put(60,0){\line(2,1){10}}
\put(60,0){\line(2,-1){10}}
\put(80,0){\makebox(0,0){+}}
\put(90,0){\line(1,0){10}}
\put(105,0){\circle{10}}
\put(110,0){\line(2,1){10}}
\put(110,0){\line(2,-1){10}}
\put(130,0){\makebox(0,0){+}}
\put(140,0){\line(1,0){10}}
\put(155,0){\circle{10}}
\put(165,0){\circle{10}}
\put(170,0){\line(2,1){10}}
\put(170,0){\line(2,-1){10}}
\put(40,-20){\makebox(0,0){+}}
\put(50,-20){\line(1,0){10}}
\put(65,-20){\circle{10}}
\put(75,-20){\circle{10}}
\put(85,-20){\circle{10}}
\put(90,-20){\line(2,1){10}}
\put(90,-20){\line(2,-1){10}}
\put(110,-20){\makebox(0,0)[l]{+ $\cdots$}}
\put(6,-8){\scriptsize $t_2$}
\put(20,-8){\scriptsize $t_1$}
\put(20,9){\vector(-1,0){10}}
\put(15,11){\scriptsize $p$}
\end{picture}}
\caption{Diagrammatic representation of the temporally extended vertex function 
$\lambda(p,t_2-t_1)$. The propagator is represented by a full line. Here is 
shown the diagrammatic expansion for the trivertex $\lambda\bar{a}a^2$. Similar 
diagrams can be 
drawn for the quadrivertex $\lambda\bar{a}^2a^2$. \label{fig:one}}
\end{figure}
We define $g_{\rm{R}}$, the dimensionless renormalized coupling (or running 
coupling constant), using the minimal subtraction scheme. That is we define 
\begin{equation}
g_{\rm{R}}=Z_{\rm{g}}\kappa^{-\epsilon}\frac{\lambda}{2\pi D} \label{eq:gR}
\end{equation}
where $\kappa$ is a normalization point and $Z_{\rm{g}}=1 + 
\sum_{i=1}^{\infty}a_i g_{\rm{R}}^i$. The $a_i$ are chosen such as to exactly 
cancel the poles of order $1/\epsilon$  appearing in $\tilde{\lambda}(p,s)$. 
From~\eref{eq:lambda}, we have
\[
\tilde{\lambda}(p,s)
=\frac{\lambda}
      {1+\kappa^{-\epsilon}\lambda[1+\epsilon A(\epsilon)]/(2\pi D\epsilon)},
\]
where $A(\epsilon)=\Or(1)$. With~\eref{eq:gR}, one finds
\[
\frac{\tilde{\lambda}(p,s)}{2\pi D}
= \frac{k^{\epsilon}g_{\rm{R}}}
       {Z_{\rm{g}} + g_{\rm{R}}/\epsilon + g_{\rm{R}}A(\epsilon)}.
\]
Accordingly to our prescription, we have to choose $a_1=-1/\epsilon$ and $a_i=0$ 
for $i>1$, giving $Z_{\rm{g}}=1-g_{\rm{R}}/\epsilon$ (exact to all orders). The 
$\beta$ function is defined by
\[
\beta(g_{\rm{R}})
\equiv \kappa\frac{\partial g_{\rm{R}}}{\partial\kappa}
= -\epsilon g_{\rm{R}} + g_{\rm{R}}^2.
\]
It is exactly quadratic in $g_{\rm{R}}$ and has a fixed point given by
$\beta(g_{\rm{R}}^{*})=0$ at $g_{\rm{R}}^{*}=\epsilon$.

The bare coupling may thus be expressed in terms of the renormalized one:
\begin{equation}
\kappa^{-\epsilon}\frac{\lambda}{2\pi D}
=\frac{g_{\rm{R}}}{1-g_{\rm{R}}/\!g_{\rm{R}}^{*}}
=g_{\rm{R}} + \frac{g_{\rm{R}}^2}{g_{\rm{R}}^{*}} + \dots \label{eq:bare}
\end{equation}
The perturbation theory can then be written as  an expansion in powers of 
$g_{\rm{R}}$.

\subsection{Renormalization group equations}

An arbitrary renormalized correlation function $G^{N,\bar{N}}_{\rm{R}} 
(\{x_i,t_i\}_1^{N+\bar{N}})$ (where the subscript R stands for renormalized) is 
related to its  bare expression~\eref{eq:defcor} through
\[
G^{N,\bar{N}}_{\rm R}(\{x_i,t_i\}_1^{N+\bar{N}};g_{\rm R},D,J,\kappa)
= G^{N,\bar{N}}(\{x_i,t_i\}_1^{N+\bar{N}};\lambda,D,J).
\]
The independence of the bare functions on the normalization scale can be 
expressed via the condition:
\[
\biggl(\kappa\frac{\partial}{\partial\kappa} 
      + \beta(g_{\rm{R}})\frac{\partial}{\partial g_{\rm{R}}} \biggr)
G^{N,\bar{N}}_{\rm{R}}(\{x_i,t_i\}_1^{N+\bar{N}};g_{\rm{R}},D,J,\kappa)=0.
\]
The formal solution (obtained by the method of characteristic) is
\begin{equation}
G^{N,\bar{N}}_{\rm{R}}(\{x_i,t_i\}_1^{N+\bar{N}};g_{\rm{R}},D,J,\kappa)
=G^{N,\bar{N}}_{\rm{R}}(\{x_i,t_i\}_1^{N+\bar{N}};g_{\rm{R}}(\rho),D,J,
 \rho\kappa) \label{eq:RGE}
\end{equation}
with
\begin{equation}
g_{\rm{R}}(\rho)
= g_{\rm{R}}^{*}
  \biggl(1 + \frac{g_{\rm{R}}^{*} - g_{\rm{R}}}{g_{\rm{R}}}\rho^\epsilon
  \biggr)^{-1}. \label{eq:rungR}
\end{equation}
Note that in the small $\rho$ limit, $g_{\rm{R}}(\rho) \to g_{\rm{R}}^{*}$.

We can implement \eref{eq:RGE} with a dimensional analysis. The dimensions of  
the different quantities, expressed in term of momentum $\kappa$ and energy $E$ 
are
\begin{eqnarray*}
[t]=E^{-1},\qquad [D]=E\kappa^{-2},\qquad [J]=E\kappa^{d}\\
\null[a]=\kappa^{d},\qquad [\bar{a}]=1,\qquad 
[G^{N,\bar{N}}_{\rm{R}}(\{x_i,t_i\}_1^{N+\bar{N}})]=\kappa^{Nd},
\end{eqnarray*}
Thus,
\begin{equation}\fl
G^{N,\bar{N}}_{\rm{R}}(\{x_i,t_i\}_1^{N+\bar{N}};g_{\rm{R}},D,J,\kappa)
= \kappa^{Nd}G^{N,\bar{N}}_{\rm{R}}(\{\kappa x_i,\kappa^2 D t_i\}_1^{N+\bar{N}}; 
  g_{\rm{R}},1,\kappa^{-d-2}J/\!D,1). \label{eq:analdim}
\end{equation}
The combination of~\eref{eq:RGE} and~\eref{eq:analdim} leads to
\begin{eqnarray}\fl
G^{N,\bar{N}}_{\rm{R}}(\{x_i,t_i\}_1^{N+\bar{N}};g_{\rm{R}},D,J,\kappa)
\nonumber \\
\lo{=} (\rho\kappa)^{Nd} G^{N,\bar{N}}_{\rm{R}}(\{\rho\kappa x_i,
       (\rho\kappa)^2 Dt_i\}_1^{N+\bar{N}};g_{\rm{R}}(\rho),1,
       (\rho\kappa)^{-d-2}J/\!D,1). \label{eq:RGEdim}
\end{eqnarray}

We can then use the following strategy to compute the correlation functions: 
first an expansion in power of $\lambda$ is established; then it is converted 
into an expansion in power of $g_{\rm{R}}$ through \eref{eq:bare}. The 
singularities in $\epsilon$ are eliminated using the renormalization scheme. 
Now, for a correctly renormalized theory, we can rewrite the $g_{\rm{R}}$ 
expansion into an $\epsilon$ expansion using~\eref{eq:RGEdim} 
and~\eref{eq:rungR}. Indeed, introducing the $\rho$ dependence 
through~\eref{eq:RGEdim} and letting $\rho\to 0$, $g_{\rm{R}}(\rho) \to 
g_{\rm{R}}^{*}$ one obtains $G^{N,\bar{N}}_{\rm{R}}$ as an expansion in power of 
$\epsilon$. 

Up to now $\rho$ is an arbitrary parameter, and several choices are possible.
For example, if we choose $\rho$ such that
\begin{equation}
(\rho\kappa)^{-d-2}\frac{J}{D}=1. \label{eq:chooserho}
\end{equation}
the limit $\rho\to 0$ becomes equivalent to $J\to 0$: when the source rate is 
vanishing small, the running coupling approaches its fixed point value. 

Another choice is
\[
(\rho\kappa)^2 Dt_1=1
\]
and the limit $\rho\to 0$ may be exchanged with $t_1\to\infty$.  However, from
\eref{eq:RGEdim}, we see that $(\rho\kappa)^{-d-2}J/\!D$ diverges.  Thus, with
this choice, one should know the behaviour of $G^{N,\bar{N}}_{\rm{R}}$ for
arbitrary large value of $J$.  However, for large values of $J$, the
$\epsilon$-expansion breaks down (see Appendix A).  Accordingly, we shall choose
$\rho$ acccording to condition~\eref{eq:chooserho} in what follows.

\subsection{Density calculation}

The density is first calculated using~\eref{eq:density}, in the framework of a
perturbation expansion in power of $\lambda$.  At the tree-level, we find out
the mean-field result and we can directly use the RG equation (no
renormalization is needed).  The first correction to the classical behaviour is
given by the one-loop contribution.  The corresponding diagram may be calculated
using the action obtained from~\eref{eq:action} by shifting the field $a$ by its 
classical value:
\begin{equation}\fl
S[\eta,\bar{\eta},J]=\int\!\d^{d}x\!\int_{0}^{t}\!\d\bar{t}\,
\biggl[\bar{\eta}
      \biggl( \frac{\partial}{\partial\bar{t}} - D\nabla^2 + 2\lambda
            a_{\rm{cl}} \biggr) \eta
      + \lambda\bar{\eta}\eta^2 
      + \lambda\bar{\eta}^2(a_{\rm{cl}}^2 + 2a_{\rm{cl}}\eta +\eta^2)
\biggl], \label{eq:etaction}
\end{equation}
where $\eta=a-a_{\rm{cl}}$ and $\bar{\eta}=\bar{a}$, with $a_{\rm{cl}}$ given
by~\eref{eq:MFdens}. We have
\[
G_{\eta\bar{\eta}}(p,t,t')
= \theta(t-t')\exp[-Dp^2(t-t')]
  \Biggl(\frac{\cosh(\sqrt{J\lambda}\,t')}{\cosh(\sqrt{J\lambda}\,t)}\Biggr)^2.
\]
Note that because of the initial condition, $G_{\eta\bar{\eta}}$ is not 
invariant under time-translation. Obtaining divergent expressions for the 
one-loop corrections, we shall renormalize them using the renormalization scheme 
developed above. We give below a summary of these results.

\subsubsection{Tree level}
Applying the RG formalism as developed above on the mean-field equation, we 
find for $J$ sufficiently small,
\begin{eqnarray}\fl
c_{\rm{R}}(t;g_{\rm{R}},D,J,\kappa) =
  \frac{1}{\sqrt{2\pi\epsilon}}\biggl(\frac{J}{D}\biggr)^{d/(d+2)} \nonumber \\
\times
  \Biggl(1 -
        \frac{2\exp[-2\sqrt{2\pi\epsilon}\,(J/\!D)^{2/(d+2)}Dt]}
             {1 + 2\exp[-2\sqrt{2\pi\epsilon}\,(J/\!D)^{2/(d+2)}Dt]}
  \Biggr)[1 + \Or(\epsilon)],
\end{eqnarray}
which is universal (independent of $g_{\rm{R}}$). This result is valid for any 
time $t$, because we only need to tune $J$ to be in the critical domain.

For small $t$ we find the expected result
\[
c_{\rm{R}}(t) = Jt + \Or(t^2)
\]
and for long time ($t\to\infty$)
\begin{equation}\fl
c_{\rm{R}}(t)
= 
\frac{1}{\sqrt{2\pi\epsilon}}\biggl(\frac{J}{D}\biggr)^{d/(d+2)}
  \{1 - 2\exp[-2\sqrt{2\pi\epsilon}\,(J/\!D)^{2/(d+2)}Dt]\}
  [1 + \Or(\epsilon)].
\end{equation}
The steady state value is thus given by
\begin{equation}
c_{\rm{R}}(\infty)
= 
\frac{1}{\sqrt{2\pi\epsilon}}\biggl(\frac{J}{D}\biggr)^{d/(d+2)}
  [1 + \Or(\epsilon)]
\end{equation}
and it is approached  exponentially in time as:
\begin{equation}
\delta c_{\rm{R}}(t) \equiv c_{\rm{R}}(t) - c_{\rm{R}}(\infty)
\propto \exp[-\gamma_{\epsilon} (J/\!D)^{2/(d+2)}Dt] \qquad (t\to\infty)
\label{eq:asympt}
\end{equation}
with $\gamma_{\epsilon}=2\sqrt{2\pi\epsilon}\,[1 + \Or(\epsilon)]$.

\subsubsection{One-loop corrections}

The diagram corresponding to the one-loop correction is given in \fref{fig:two}; 
its analytic expression is
\begin{figure}[bt]
\setlength{\unitlength}{1mm}
\centerline{
\begin{picture}(17,20)(-1,-10)
\put(0,0.25){\line(1,0){5}}
\put(0,-0.25){\line(1,0){5}}
\put(10,0){\circle{10}}
\put(10,0){\circle{9}}
\put(14.75,0){\circle*{1.5}}
\put(-0.5,7){$t$}
\put(4.5,7){$t_2$}
\put(14.5,7){$t_1$}
\put(10,4.5){\makebox(0,0){\scriptsize $<$}}
\put(10,3.5){\makebox(0,0)[t]{\scriptsize $p$}}
\put(10,-4.5){\makebox(0,0){\scriptsize $<$}}
\put(10,-5.5){\makebox(0,0)[t]{\scriptsize $-p$}}
\end{picture}}
\caption{One-loop diagram for the density, using 
action~\protect\eref{eq:etaction}. The double line stands for the free 
propagator $G_{\eta\bar{\eta}}$ and the dot for the vertex 
$\lambda a_{\rm{cl}}^2 \bar{\eta}^2$. \label{fig:two}}
\end{figure}
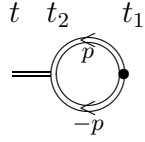
\begin{eqnarray*}\fl
c^{(1)}(t)
= \frac{2J\lambda}{\cosh^2(\sqrt{J\lambda}\,t)}
  \int_0^t\!\d t_2\!\int_0^{t_2}\!\d t_1 \int 
\frac{\d^{d}p}{(2\pi)^{d}}
  \exp[-2Dp^2(t_2-t_1)] \\
\times
  \frac{\cosh^2(\sqrt{J\lambda}\,t_1)\sinh^2(\sqrt{J\lambda}\,t_1)}
       {\cosh^2(\sqrt{J\lambda}\,t_2)}.
\end{eqnarray*}
The integral over the momentum gives the factor $[8\pi D(t_2-t_1)]^{-d/2}$. We 
thus obtain
\begin{eqnarray*}\fl
c^{(1)}(t)
= \frac{2J\lambda}{(8\pi D)^{-d/2}}
  \frac{t^{1+\epsilon/2}}{\cosh^2(\sqrt{J\lambda}\,t)}
  \int_0^1\!\d x_2\!\int_0^{1}\!\d x_1\,x_1^{-1+\epsilon/2} \\
\times
  \frac{\cosh^2[\sqrt{J\lambda}\,x_2(1-x_1)]
        \sinh^2[\sqrt{J\lambda}\,x_2(1-x_1)]}
       {\cosh^2(\sqrt{J\lambda}\,x_2)}.
\end{eqnarray*}
In the framework of an $\epsilon$-expansion, we eventually find (treating 
$x_1^{-1+\epsilon/2}$ as a generalized function~\cite{Russes}, see Appendix B):
\[\fl
c^{(1)}(t)
= \frac{2J\lambda}{(8\pi D)^{-d/2}}
  \frac{t^{1+\epsilon/2}}{\cosh^2(\sqrt{J\lambda}\,t)}
  \biggl(\frac{2}{\epsilon}
        \frac{\sinh(2\sqrt{J\lambda}\,t) - 2\sqrt{J\lambda}\,t}
             {2\sqrt{J\lambda}\,t}
        + \frac{\phi(\sqrt{J\lambda}\,t)}{\sqrt{J\lambda}\,t} + \Or(\epsilon)
  \biggr),
\]
where
\begin{eqnarray*}\fl
\phi(\xi)=\xi \int_0^1\!\d x_2\,\ln x_2\,\sinh^2(\xi x_2) \\
+ \xi \int_0^1 \frac{\d x_1}{x_1} \int_0^1\!\d x_2
  \biggl(\frac{\cosh^2[\xi x_2(1-x_1)]\sinh^2[\xi x_2(1-x_1)]}
              {\cosh^2(\xi x_2)} -\sinh^2(\xi x_2)
  \biggr).
\end{eqnarray*}
Putting together the mean-field result and the one-loop correction, we verify 
that the density is indeed divergence free, and using the RG 
equation~\eref{eq:RGEdim} we find, for small values of $J$
\begin{eqnarray}\fl
c_{\rm{R}}(t;g_{\rm{R}},D,J,\kappa)
= \frac{1}{\sqrt{2\pi\epsilon}}\biggl(\frac{J}{D}\biggr)^{d/(d+2)}
  \Biggl\{ \tanh[\sqrt{2\pi\epsilon}\,(J/\!D)^{2/(d+2)}Dt] \nonumber \\
+       \frac{\epsilon}{2}
        \Biggl[\frac{\phi[\sqrt{2\pi\epsilon}\,(J/\!D)^{2/(d+2)}Dt]}
                    {\cosh^2[\sqrt{2\pi\epsilon}\,(J/\!D)^{2/(d+2)}Dt]}
              + \frac{1}{2}\ln[8\pi(J/\!D)^{2/(d+2)}Dt] \nonumber \\
\times        \Biggl(\tanh[\sqrt{2\pi\epsilon}\,(J/\!D)^{2/(d+2)}Dt]
                    - \frac{\sqrt{2\pi\epsilon}\,(J/\!D)^{2/(d+2)}Dt}
                           {\cosh^2[\sqrt{2\pi\epsilon}\,(J/\!D)^{2/(d+2)}Dt]}
              \Biggr)
       \Biggr] + \Or(\epsilon^2)
  \Biggr\} \nonumber \\
\lo=
  \biggl(\frac{J}{D}\biggr)^{d/(d+2)}{\cal F}[(J/\!D)^{2/(d+2)}Dt].
  \label{eq:scalef}
\end{eqnarray}
We immediately identify the scaling function $\cal F$ defined 
 in~\eref{eq:scaling}. Taking the limit $t\to\infty$ we find
\begin{equation}\fl
c_{\rm{R}}(\infty;g_{\rm{R}},D,J,\kappa)
= \frac{1}{\sqrt{2\pi\epsilon}}\biggl( \frac{J}{D}\biggr)^{d/(d+2)}
  \biggl\{ 1 - \frac{\epsilon}{4}
        \biggl[ \gamma_{\rm{E}} -\ln\biggl(\frac{2\pi}{\epsilon}\biggr) \biggr]
        + \Or(\epsilon^2)
  \biggr\},
\end{equation}
where $\gamma_{\rm{E}}$ is the Euler constant ($\gamma_{\rm{E}} \simeq 0.5772$). 
Unfortunately, due to the complicated form of $\phi$, we are unable to give a 
more compact form for the asymptotics of $\cal F$. Notice that the empty initial 
condition implies that $\lim_{J\to 0} c_{\rm{R}}(t)=0$.

Let us consider two particular cases of interest. The first one is the case  
$\epsilon=1$ ($d=1$). Then the steady-state density is
\begin{equation}
c_{\rm{R}}(\infty)
= \frac{1}{\sqrt{2\pi}} \biggl( \frac{J}{D}\biggr)^{1/3}
   [1 - \hbox{$\frac{1}{4}$}(\gamma_{\rm{E}} -\ln 2\pi) + \cdots] 
\label{eq:RGdens}
\end{equation}
and it is asymptotically approached  (at the tree-level) as:
\begin{equation}
\delta c_{\rm{R}}(t) \propto \exp[-\sqrt{8\pi}(J/\!D)^{2/3}Dt]. \label{eq:RGapp}
\end{equation}
We shall compare the accuracy of these expressions with exact results in the 
next section. Note however that, in principle, nothing ensures us that the terms 
we neglected are small.

The second case is  $\epsilon=0$ ($d=2$). The running coupling is given by
\[
g_{\rm{R}}(\rho) = \frac{g_{\rm{R}}}{1-g_{\rm{R}}\ln\rho},
\]
which goes to $-(\ln\rho)^{-1}$ when $\rho\to 0$. By replacing this expression 
in our previous formula, we find for the steady-state density
\begin{equation}
c_{\rm{R}}(\infty)
= \biggl( \frac{2J}{\pi D}\biggr)^{1/2}
  \Bigl[\Bigl(\ln(J/\!\kappa^4 D)\Bigr)^{-1/2}
       + \Or\Bigl([\ln(J/\!\kappa^4 D)]^{-1}\Bigr)
  \Bigr]
\end{equation}
and
\begin{equation}
\delta c_{\rm{R}}(t)
\propto \exp[-\sqrt{8\pi}(J/\!D)^{1/2}\ln(J/\!\kappa^4 D)Dt].
\end{equation}
as anticipated, logarithmic corrections to the mean-field result are obtained.

\section{Exact results in one dimension}

A large amount of work has been done to solve exactly one-species diffusion
reaction models in one dimension (see for example~\cite{exact}).  In particular,
the diffusion-annihilation and the diffusion-coagulation reactions have been
considered with an input of particles.  In the diffusion-annihilation case,
R\'acz~\cite{RaczTwo} obtained the steady-state concentration by mapping its
model to the kinetic Glauber-Ising model~\cite{Glauber}.  In 1988, Doering and
ben-Avraham~\cite{Dben} calculated exactly the time dependent concentration for
a simple diffusion-coagulation model, using the interparticle distribution
function.  Since, their method has been generalized to other
diffusion-coagulation processes (see for example~\cite{Bur}) and in particular
the steady-state concentration has been obtained~\cite{Doeben} for the
diffusion-coagulation with an input source of particle.

In this section we aim at testing the validity of the RG predictions for $d=1$.
For this purpose we shall extend Doering's and ben-Avraham's results and compute
the time-dependent concentration.

We consider an infinite chain (our one-dimensional space) initially empty, and
we allow particles to appear randomly at rate $J$ (per unit time and per unit
length).  Thus initially
\[
\frac{\d c(t)}{\d t}\bigg|_{t=0}=J
\]
where $c(t)$ is as before the particle concentration.  The particles diffuse on
the line (with a diffusion constant $D$) and when two particles meet, they
instantaneously coagulate ($A+A\to A$).  Note that this model is the same as the
pure diffusion-coagulation process of section 2, but with an infinite reaction
rate $\lambda$.  Of course, for such a reaction rate, the perturbation expansion
in power of $\lambda$ is meaningless and one may argue that the two models are
not equivalent.  However by examining the relation between the renormalized
coupling $g_{\rm{R}}$ and $\lambda$, we see (from~\eref{eq:bare}) that:
\[
g_{\rm{R}}
= g_{\rm{R}}^{*} (1 + 2\pi g_{\rm{R}}^{*}\kappa^{\epsilon}D/\!\lambda)^{-1};
\]
and thus, when $\lambda$ is infinite, $g_{\rm{R}}=g_{\rm{R}}^{*}$.  We argue
that the infinite reaction rate limit may be obtained by taking the fixed point
coupling limit ($g_{\rm{R}}\to g_{\rm{R}}^{*}$), that is to say by taking the
$\lambda\to\infty$ limit {\it after} having performed the path integrals.  This
will be confirmed in the following, at least in one dimension.

In a one-dimensional space, the particles concentration can be related to the 
probability $E(x,t)$ that an interval of length $x\geq 0$ is empty at time $t$, 
via
\[
c(t)=-\frac{\partial E(x,t)}{\partial x}\bigg|_{x=0}.
\]
As shown by Doering and ben-Avraham~\cite{Doeben}, $E(x,t)$ has the advantage to 
obey a closed equation of evolution, namely:
\begin{equation}
\frac{\partial E}{\partial t} = 2D \frac{\partial^2 E}{\partial x^2} - JxE,
\label{eq:evolE}
\end{equation}
with the two conditions
\begin{equation}
E(0,t)=1,\qquad E(\infty,t)=0. \label{eq:condE}
\end{equation}
From this equation, one can immediately obtain the steady-state, by setting the 
left hand side to zero. One then recognizes the Airy equation, whose solution is 
(taking into account conditions~\eref{eq:condE})
\[
E(x,\infty)=\frac{{\rm Ai}\Bigl((J/2D)^{1/3}x\Bigr)}{{\rm Ai}(0)},
\]
where ${\rm Ai}(z)$ is the Airy's function (see~\cite{Abramo}). As a 
consequence, the asymptotic concentration reads
\begin{equation}
c(\infty) = -\frac{{\rm Ai}'(0)}{{\rm Ai}(0)}\biggl(\frac{J}{2D}\biggr)^{1/3}
\label{eq:asymconc}
\end{equation}
${\rm Ai}'(z)$ is the first derivative of ${\rm Ai}(z)$. Note that ${\rm 
Ai}'(0)<0$.

Before comparing~\eref{eq:asymconc} with the RG results, we shall compute the 
time dependent part of the concentration. For this purpose, we shall 
solve~\eref{eq:evolE} using the Laplace transform $\tilde{E}(x,s)$ 
defined by $\tilde{E}(x,s) = \int_0^t\!\d t\,e^{-st}E(x,t)$. \Eref{eq:evolE} 
becomes
\begin{equation}
2D \frac{\partial^2 \tilde{E}(x,s)}{\partial x^2} - (Jx+s)\tilde{E}(x,s)
+ E(x,0) = 0, \label{eq:evolEtilde}
\end{equation}
where $E(x,0)$ is the initial condition (for an empty system, $E(x,0)=1$). 
\Eref{eq:evolEtilde} is an inhomogeneous second order ordinary differential 
equation. Its general solution is the sum of the homogeneous solution 
$\tilde{E}_{\rm{h}}(x,s)$ and a particular solution. The homogeneous solution is
\[
\tilde{E}_{\rm{h}}(x,s)
= \biggl(\frac{J}{2D}\biggr)^{-2/3}
  [ \alpha_1(s){\rm Ai}(z) + \alpha_2(s){\rm Bi}(z)],
\]
where $z=(J/2D)^{-2/3}(Jx+s)/2D$, $\alpha_1(s)$ and $\alpha_2(s)$ are two 
unknown functions of $s$ and ${\rm Bi}(z)$ is the second Airy's 
function~\cite{Abramo}.

Writing $\tilde{E}(x,s)=(J/2D)^{-2/3}A(z,s)$, \eref{eq:evolEtilde} becomes
\[
\frac{\partial^2 A(z,s)}{\partial z^2} - zA(z,s) = -\frac{1}{2D},
\]
for which a well known solution is
\[
A(z,s) = -\frac{\pi}{2D}\int_0^z\!\d v\,
         [{\rm Ai}(v){\rm Bi}(z) - {\rm Ai}(z){\rm Bi}(v)].
\]
The two boundary conditions~\eref{eq:condE} permit us to determine the two 
unknown functions $\alpha_1(s)$ and $\alpha_2(s)$. We eventually find for the 
general solution of~\eref{eq:evolEtilde}
\begin{eqnarray}\fl
\tilde{E}(x,s)
= \frac{\pi}{2D}\biggl(\frac{J}{2D}\biggr)^{-2/3}
  \biggl\{{\rm Ai}(\xi+\sigma)\int_{\sigma}^{\xi+\sigma}\!\d v\,{\rm Bi}(v)
        + {\rm Bi}(\xi+\sigma)
        \biggl[\frac{1}{3} - \int_{0}^{\xi+\sigma}\!\d v\,{\rm Ai}(v)\biggr]
  \nonumber \\
+       \frac{{\rm Ai}(\xi+\sigma)}{{\rm Ai}(\sigma)}
        \biggl[\frac{1}{\pi\sigma} + {\rm Bi}(\sigma)
              \biggl(\int_0^{\sigma}\!\d v\,{\rm Ai}(v) - \frac{1}{3}\biggr)
        \biggr]
  \biggr\}, \label{eq:Esol}
\end{eqnarray}
where $\xi=(J/2D)^{1/3}x$ and $\sigma=(J/2D)^{-2/3}s/2D$.

The probability $E(x,t)$ is then obtained by Laplace-inverting~\eref{eq:Esol}. 
For $t>0$, we only have to care for the poles of $\tilde{E}(x,s)$. They are 
located at $\sigma=0$ and at $\sigma=a_n$, $n=1,2,3,\ldots$, where $a_n$ is the 
$n$-th zero of ${\rm Ai}(x)$ ($a_n<0$). We finally obtain 
\[\fl
E(x,t)
= \frac{{\rm Ai}(\xi)}{{\rm Ai}(0)} + \sum_{n=1}^{\infty}
  \frac{{\rm Ai}(\xi+a_n)}{{\rm Ai}'(a_n)}
  \biggl[\frac{1}{a_n} + \pi{\rm Bi}(a_n)
         \biggl(\int_0^{a_n}\!\d v\,{\rm Ai}(v) - \frac{1}{3}\biggr)
  \biggr] \e^{-|a_n|\tau},
\]
with $\tau=2Dt(J/2D)^{2/3}$ and
\begin{equation}\fl
c(t)
= \biggl(\frac{J}{2D}\biggr)^{1/3}
  \biggl\{-\frac{{\rm Ai}'(0)}{{\rm Ai}(0)} - \sum_{n=1}^{\infty}
        \biggl[\frac{1}{a_n} + \pi{\rm Bi}(a_n)
              \biggl(\int_0^{a_n}\!\d v\,{\rm Ai}(v) - \frac{1}{3}\biggr)
        \biggr] \e^{-|a_n|\tau}
  \biggr\}. \label{eq:exdens}
\end{equation}

We are now in position to compare these results with the RG results for 
$\epsilon=1$. For the steady-state, the RG gives (up to the one-loop 
corrections)~\eref{eq:RGdens}
\begin{equation}
c_{\rm R}(\infty) \simeq 0.53 \biggl(\frac{J}{D}\biggr)^{1/3}
\end{equation}
(for small $J$) while (putting $t=\infty$ inside~\eref{eq:exdens}) the exact 
solution gives: 
\begin{equation}
c(\infty)
= -\frac{{\rm Ai}'(0)}{{\rm Ai}(0)}\biggl(\frac{J}{2D}\biggr)^{1/3}
\simeq 0.58 \biggl(\frac{J}{D}\biggr)^{1/3}
\end{equation}
(for arbitrary $J$). Surprisingly, the difference is only of the order of 10\%.

The comparison for the approach to the steady-state is less convincing, mainly 
due to the fact that we do not know the one-loop corrections. The RG gives, from 
equation~\eref{eq:RGapp}
\begin{eqnarray}
\delta c_{\rm R}(t)
&= -\Bigl[1 - \hbox{$\frac{1}{4}$}(\gamma_{\rm{E}} -\ln 2\pi)\Bigr]
   \biggl(\frac{2}{\pi}\biggr)^{1/2}
   \biggl(\frac{J}{D}\biggr)^{1/3} \exp[-\sqrt{8\pi} (J/\!D)^{2/3}Dt]
   \nonumber \\
&\simeq
   -1.06 \biggl(\frac{J}{D}\biggr)^{1/3} \exp[-5.01 (J/\!D)^{2/3}Dt],
\end{eqnarray}
whereas the exact result is
\begin{eqnarray}
\delta c(t)
&= \biggl[\frac{1}{a_1} + \pi{\rm Bi}(a_1)
         \biggl(\int_0^{a_1}\!\d v\,{\rm Ai}(v) - \frac{1}{3}\biggr)
   \biggr] \biggl(\frac{J}{2D}\biggr)^{1/3} \e^{-|a_1|\tau} \nonumber \\
&\simeq
   -1.10 \biggl(\frac{J}{D}\biggr)^{1/3} \exp[-2.95 (J/\!D)^{2/3}Dt]
\end{eqnarray}
(with $a_1\simeq -2.33$). Both amplitudes in front of the exponential are in 
good agreement (because we used the one-loop result of the steady-state).
However, the tree-level amplitude into the exponential is quite different (of 
almost 70\%) from the one given by the exact theory. The inclusion of the 
one-loop correction should lead to a better agreement.

Note that the exact results are valid without any restriction on $J$, on the
contrary to the RG results, which apply only for small $J$.  This restriction
was introduced to ensure that $g_{\rm R}$ to be inside the critical domain ({\it
i.e.}  $g_{\rm R}$ near $g_{\rm R}^{*}$).  In view of the predictions of the
exact theory, it turns out that this restriction over $J$ is unnecessary.  The
coupling stays inside the critical domain for any value of $J$.  This justifies
why when $\lambda\to\infty$, $g_{\rm R}\to g_{\rm R}^{*}$.

\section{Discussion and conclusive remarks}

We have shown that the RG is a suitable formalism to compute the density in a
wide class of diffusion-reaction model, with an input of particle.  In
particular, we easily calculated the critical exponents in arbitrary dimension.
However, the computation of the universal scaling function is generally much
more difficult (see for example~\eref{eq:scalef}) and can only be achieved
within a power expansion in $\epsilon=2-d$.  However, the comparison of the
first order results for $\epsilon=1$ are not to far from the results obtained
exactly in one dimension by a different approach.

One of the main interest of the RG method is that, contrary to most
unidimensional exact approach, the RG approach is not restricted to the
computation of the particle density but higher correlation functions can also be
calculated (although the computation can become quite involved,
see~\cite{Lee}).

Another advantage of RG approach, is the fact that the properties of a whole
class of model can be simply related.  This allows us using~\eref{eq:densrel},
and knowing the particle density for the diffusion-coagulation model, to obtain
the density of any mixed annihilation-coagulation process.  In particular, for
the pure diffusion-annihilation model ($\alpha=2$) in one dimension, one
recovers the steady-state density previously calculated by
R\'acz~\cite{RaczTwo}.

The present work can be generalized in several directions.  Generalization to
$mA\to\emptyset$ ($m>2$) reactions with a source of particles is also
straightforward.  Indeed, following Lee, it turns out that apart the lowering of
the critical dimension to $d_{\rm u}=2/(m-1)$, only minor changes occur (for
example the amplitudes become $m$-dependent).  This is due to the fact that in
our renormalization scheme only the fixed point $g_{\rm R}^{*}$ of the coupling
is modified.  The structure of the equations remain unchanged and in particular,
the RG equation~\eref{eq:RGEdim} still holds.  As a consequence, the critical
exponents below $d_{\rm u}$ are the same than for $m=2$.  The steady-state
concentration can be written (for $d<d_{\rm u}$) as:
\[
c_{m,{\rm R}}(\infty)=A_m \biggl( \frac{J}{D} \biggr)^{d/(d+2)},
\]
and the universal amplitude $A_m$ can be computed within an
$\epsilon$-expansion.  The approach to the steady-state is still exponential
with $k$-dependent amplitudes.  Extension to the reactions $mA\to lA$ with $l<m$
($m>2$) is also possible (see~\cite{Lee}).

Another natural extension of this work is to consider the possibility of
reversible reaction (for example $A+A\rightleftharpoons A$).  For such system a
completely different physics is expected.  Indeed, a quick investigation of the
associated action shows that the upper critical dimension is no longer 2 but 4.
In addition, it appears that a wavefunction renormalization is needed, giving
rise to anomalous dimensions.  The computation of scaling functions for this problem is currently under investigation.

{\appendix
\section{}

We aim to show that in the limit $J \to \infty$, the $\epsilon$-expansion breaks down.
For this purpose, let us consider the action~\eref{eq:etaction} obtained by the
shift of the field $a$ by its classical value.  The shift permits us to suppress
the source term in the action; in other words, shifting the field is equivalent
to perform the infinite sum generated by the source term.  One could then think
that any quantity can be calculated for arbitrary $J$.  This is however not true
for the following reason.  To calculate a given quantity, we expand the action
with respect to the $\lambda\bar{\eta}^2(a_{\rm{cl}}^2 +
2a_{\rm{cl}}\eta+\eta^2)$ term.  We shall then obtain a power expansion in term
of $\lambda$ with coefficient proportional to $a_{\rm{cl}}^2$.  But for large
$J$, $a_{\rm{cl}}$ tends to $(J/\!\lambda)^{1/2}$.  The expansion in $\lambda$
is partly replaced by an expansion in $J$, which does not lead to an
$\epsilon$-expansion.  The only possibility to revert this would be to treat
non-perturbatively the action, which is out of question.

\section{}

The problem is to compute the $\epsilon$-expansion of
\begin{equation}
\int_0^1\d x\,x^{-1+\epsilon}f(x). \label{eq:genfunc}
\end{equation}
Putting naively $x^{-1+\epsilon}=x^{-1}[1 + \epsilon
\ln x + \Or(\epsilon^2)]$ in~\eref{eq:genfunc} clearly fails, because if 
$f(0)\not=0$, $\int_0^1\d x\,x^{-1}f(x)$ diverges. One way to avoid this 
problem, is to treat $x^{-1+\epsilon}$ as a generalized function.
We shall not give here a detailed discussion of the 
generalized functions (see~\cite{Russes} for an introduction). We only quote the result, namely:
\begin{eqnarray*}\fl
\int_0^1\d x\,x^{-1+\epsilon}f(x)
= \frac{1}{\epsilon}f(0) + \int_0^1\d x\,x^{-1}[f(x)-f(0)] \\
+ \epsilon\int_0^1\d x\,x^{-1}\ln x [f(x)-f(0)] + \Or(\epsilon^2).
\end{eqnarray*}
Remark that there is a  pole of order $1/\epsilon$.

}

\end{document}